\documentstyle[12pt]{article}

\begin{document}
\baselineskip 24pt
\newcommand{\numero}{SHEP 95-33}   %Enter SHEP preprint number

\newcommand{\titre}{Reconstructed CKM Matrices }
\newcommand{\auteura}{K.J. Barnes}
\newcommand{\auteurb}{O.J. Senior}
\newcommand{\auteurc}{N.D. Virgo}
\newcommand{\place}{Department of Physics,\\
University of Southampton\\
Southampton SO17 4JU \\ U.K. }
\newcommand{\beq}{\begin{equation}}
\newcommand{\eeq}{\end{equation}}

\newcommand{\abstrait}{We construct quark mixing matrices
within a group theoretic
framework which is easily applicable to any number of
generations.
Familiar cases are retrieved and related, and it is hoped that our
viewpoint may have advantages
both phenomenologically and for constructing underlying mass
matrix schemes.}
\begin{titlepage}
\hfill \numero  \\

\vspace{.5in}
\begin{center}
{\large{\bf \titre }}
\bigskip \\ by \bigskip \\ \auteura \bigskip \\  \auteurb \\
  \bigskip  and \bigskip \\ \auteurc
    \bigskip \\ \place \bigskip \\

\vspace{.9 in}
{\bf Abstract}
\end{center}
\abstrait
 \bigskip \\
\end{titlepage}

At a recent meeting in Meribel one of the authors (K.J.B.) was
struck by
a particular presentation by B. Kayser
\cite{{Kayser1},{Kayser2}} and the reaction
of many participants in the audience.
The topic was that of describing the $CKM$ matrix
 \cite{{K&M},{NC},{RevPP},{C&K},{B&C}}
in terms of 4 phases of the unitarity triangles \cite{RevPPref},
and the connection with CP violation.
Although the audience might well be considered ``expert'', there
was a marked
resistance to consider seriously anything other than formulations
(presumably many and varied) with which participants were
already working,
and the questions and comments revealed clear
misunderstandings of
other schemes and parameterizations.
This brief note is an attempt to encourage a wider appreciation
of the parameterizations of the $CKM$ matrix and the
connections between
them in a very simple manner.
It is directly applicable to larger numbers of generations of
quarks should this
turn out to be forced by physics in the future.

In the Standard Model with $n$ generations the
Cabibbo-Kobayashi-Maskawa ($CKM$) matrix
appears as an $n \times n$ unitary matrix, $( V_n)_\alpha^\beta$,
mixing the $n$ left handed lower weak isospin quarks, $D_\beta$,
to reflect the change of
basis of the quarks from current eigenstates to mass eigenstates.
The charged current then couples these to the adjoint of the $n$
left-handed higher weak isospin
mass eigenstates $\overline{U}^\alpha$.
This notation is chosen to emphasise the group theoretic $U(n)$
nature of the structure,
although no physical symmetry need be ascribed to this group.
In particular, it should be noted that this group should not be
confused either with the weak
$SU(2) \times U(1)$ group, nor with the earlier hadronic
organising $SU(3)$ group of
Gell-Mann or its extensions to $SU(4)$, etc.
Nevertheless, from the point of view of parameterising the $CKM$
matrix the Gell-Mann \cite{GM}
notation will prove to be very convenient.
Thus the $D_\alpha$ are assigned to the $n$ dimensional
fundamental representation,
as indeed are the $U_\alpha$ so that $\bar{U}^\alpha$ are in the
conjugate multiplet.
Thus $(V_n)_\alpha^\beta$ may be expanded in any general
unitary form
(such as exponential) in terms of the adjoint multiplet represented
by the
$(n^2-1)$ Gell-Mann $(n \times n)$ hermitian
$\underline{\lambda}$ matrices and the unit.
This makes clear that $V_n$ contains $n^2$ real parameters, but
since the relative
phases of the elements of $D_\alpha$ and $\overline{U}^\beta$
can be independently picked
$(2n - 1)$ of the parameters can be removed (the overall phase
is irrelevant) and
$(n - 1)^2$ independent real parameters suffice.
We shall see later in particular cases how these give rise in
general to both real and
complex elements of $V$, and are frequently interpreted as
``mixing angles'' and
``complex  phases''.
These latter give rise to the possibility of $CP$ violation if $n >
2$.
In other interpretations the parameters are viewed as ``phases'' of
Unitarity triangles.
The physics is, of course, independent of these interpretations
although they may be useful for
visualisation of the phenomenology, or lead to intuitions as to the
underlying mass mechanisms,
and should not be undervalued.

We will now show how the $\lambda$ matrix framework is
directly relevant to implementing the
phase freedoms mentioned above, and actually constructing
useful parameterizations of $V_n$.
It is convenient to recall that the Gell-Mann representation of the
$\lambda$ matrices can be built
up inductively as $n$ increases.
Since the rank of $SU(n)$ is $(n - 1)$, there are $(n - 1)$
diagonal traceless matrices designated
$\lambda_{k (k + 2)}$, with $k = 1$ to $n - 1$,
where $\lambda_{k (k + 1)}$ has entries $ \sqrt{ \frac{2}{k ( k
- 1)}}$
down the first $(k - 1)$ diagonal places, $ (-) \sqrt{ \frac{2 ( k
- 1)}{k}}$ in the next
diagonal place, and zeros in all other places, so that the trace of
its square is two.

The off diagonal matrices follow the pattern of the Pauli matrices
both in terms of ordering
sequence and entries.
Thus $\lambda_1$ has a $1$ in the first row and second column,
and $\lambda_2$ has a $-i$ in
the same place,
all other entries being zero except for the complex conjugate
entries to the forementioned in the
transposed matrix position to ensure hermiticity.
The traceless property is obvious, as is the continuation of the
normalization that the trace of each
square is two.
Clearly the number of off diagonal matrices thus constructed to
be listed as $\lambda_i$, with $(k
- 1)^2 \leq  i \leq k^2 - 2$, is $2(k - 1)$,
since the entries $1$ and $-i$ are placed sequentially in the rows
of the $k^{th}$ column starting
in the
first row and going down to the row $(k - 1)$ as $k$ increases
through the range specified.
It follows that the number of off diagonal matrices in all is $n( n
- 1)$, and these together with
the $(n-1)$ diagonal matrices yield the full basis of $n^2 - 1$
traceless hermitian matrices.

We are now in a position to make a preliminary discussion of
what is meant by a
parameterization of $V_n$, at this stage not considering the
removal of phases
and so dealing with $n^2$ real parameters.
Obviously, since we have a basis of $n^2$ hermitian matrices,
comprising the $\lambda_i$ and
the unit, a familiar unitary parameterization is available in the
form
\beq
V_n = \exp \left[ \frac{-i}{2} ( \theta_i \lambda^i + \chi 1 )
\right]
\label{1}
\eeq
where the $n^2$ parameters, $ \theta_i$ and $\chi$, are all real.
Of course, many other unitary constructions are possible, and
these include products of several
factors each of which are unitary (usually exponential).
Two conditions must be observed.
The parameters are essential in the technical sense, so that there
must be $n^2$ of them.
If the expansion of a given parameterization for small parameters
coincides with the
expansion of equation (\ref{1}),
then these may be viewed as equivalent.
[We ignore parameterizations at large values of the angles.]
But this is not the only possibility.
It is also acceptable if the expansion involves only a subset of the
matrices, but these yield the full
set of matrices under repeated commutation.
This will include important known cases of parameterization as
will be demonstrated shortly in
particular examples,
but the idea may already be familiar to the reader through the
Euler angle specification of
rotations in three dimensions.
There the three rotations are not about three independent
orthogonal axes, but two are about a
single axis with the third separating these two being about a
second axis.
The expansion for small angles only contains two independent
infinitesimal generators
(although three parameters are used) but these commute to
produce the third infinitesimal
generator as the $A_1$ algebra of the $SO(3)$ group closes.

We now turn to the main tasks of removing the phases in such a
manner that the resulting final
form of $V_n$ is perspicuously exhibited.
The precise initial specification of $V_n$ is intimately related to
the way in which the phases are
treated in our prescription.
Two technical points arise, and as only the first is needed in the
simplest $n=2$ case, we again
turn to treating the problem iteratively.

In the $n=2$ case there are phase freedoms

\beq
D =
\left( \begin{array}{c}d \\ s \end{array} \right)
\rightarrow  \exp \left( - i \xi T_3 \right) D \; ,
\label{2}
\eeq
and

\beq
\overline{U} = ( \overline{u} \overline{c} ) \rightarrow
\overline{U} \exp ( i \omega T_3 ) \exp
\left( \frac{i }{2} \chi 1  \right) \; ,
\label{3}
\eeq
where we have reverted to the usual Pauli matrix notation $\tau^i
( i = 1,2,3)$ for the
$\lambda^i$, $T_i = \tau_i/2$ and $\xi, \omega$ and
$\chi$ are real.
Notice that, as the overall phase is physically irrelevant, there is
no $\chi$ term in equation
(\ref{2}) corresponding to the one in equation  (\ref{3}).
Indeed the overall phase is always trivial to treat, and we have
here denoted it by $\chi$ to
emphasize that it will immediately eliminate the corresponding
phase in equation  (\ref{1}).
It should now be clear that equation  (\ref{1}) is not the most
convenient starting specification
for $V_2$.
Clearly a product form
\beq
V_2 = \exp \left( \frac{-i}{2} \chi 1 \right) \exp ( - i \rho T_3 )
exp ( - i \theta_A T_A )
\label{4}
\eeq
where $A = 1,2$ , gives an immediate improvement, exposing the
diagonal matrices on the left of the structure.
But this can be further exposed by considering the form of the
right hand term, $K_2$ say.
Observe that $\theta_A$ can be regarded as the components of
a two dimensional vector rotated
by the $U_1$ factor generated by $T_3$ in our $SU_2$.
Thus, the form of $K_2$ can be re-expressed as
\beq
K_2 ( \varepsilon, \theta, T) = \exp ( - i \varepsilon T_3)  \exp (
2 i \theta T_2) \exp ( i \varepsilon
T_3)
\label{5}
\eeq
where the parameters $\varepsilon$ and $\theta$ replace the
original $\theta_A$.
[The connection between the two sets of parameters is trivial to
establish, but is not required
here.]
Substituting this form back into equation  (\ref{4}) reveals
\beq
V_2 = \exp \left( \frac{ - i \chi}{2} \right) \exp ( - i [ \varepsilon
+ \rho ] T_3) \exp ( 2 i \theta T_2
) \exp ( i \varepsilon T_3 ) ,
\label{6}
\eeq
and comparison with equations (\ref{2}) and (\ref{3}) shows that
the phase changes specified by $\chi$,
$\omega = \rho$, and $\xi = \varepsilon$, produce
\beq
V_2 = \exp ( 2 i \theta T_2)
\label{7}
\eeq
as our final one parameter description of $V_2$.
The concrete matrix form of $V_2$ is now
\beq
V_2 = \left[
\begin{array}{rr}
\cos \theta & \sin \theta \\
- \sin \theta & \cos \theta
\end{array}
\right]
\label{8}
\eeq
revealing that $\theta$ is the well known Cabibbo \cite{NC}
angle.

Now we turn to the $n=3$ case which is currently of most
physical interest.
To make optimal use of the analysis used so conveniently in the
$n=2$ case above,
we propose to exploit the $SU(2) \times U(1)$ subgroup
structure of $SU(3)$ so extensively
developed some 30 years ago by Lipkin and collaborators
\cite{LL&M}, and the version we
present differs from the presentation of Carruthers
\cite{Carruthers} only by trivial signs
specifically introduced for our present interests. With the notation
introduced earlier, the $SU_2$
generators are
\beq
T_i = \frac{\lambda_i}{2}
\label{9}
\eeq
exactly as before but extended by a third row and column of
zeros, and we introduce
\beq
T = \frac{1}{2 \sqrt{3}} \lambda_8
\label{10}
\eeq
as the generator of the $U_1$ which commutes to zero with each
of the $T_i$.
It must be emphasised, of course, that the generalisation of $D$
is now to a three component
column with descending entries $d,s$ and $b$.
Similarly the generalization of $ \overline{U}$ is to
$(\overline{u} \; \overline{c} \; \overline{t} )$.
We then introduce another set of $SU_2$ generators by
\beq
U_1 = \frac{\lambda_6}{2}, \; U_2 = \frac{\lambda_7}{2}, \;
U_3
= \frac{\sqrt{3}}{4} \lambda_8
-
\frac{1}{4} \lambda_3
\label{11}
\eeq
and an associated $U_1$ generator by
\beq
U = \frac{-1}{4 \sqrt{3}} \lambda_8 - \frac{1}{4} \lambda_3
\label{12}
\eeq
by copying the structure of $T_i$ and $T$ which distinguished the
third row and column but now
distinguishing the first row and column.
Yet again we define generators of a third $SU_2$ by
\beq
V_1 = \frac{\lambda_4}{2}, \;
V_2 = \frac{- \lambda_5}{2},
\;V_3 = - \frac{\sqrt{3}}{4} \lambda_8
-
\frac{1}{4} \lambda_3
\label{13}
\eeq
and  an associated $U_1$ generator by
\beq
V = \frac{-1}{4 \sqrt{3}} \lambda_8 + \frac{1}{4} \lambda_3
\label{14}
\eeq
by distinguishing this time the second row and column, and
judiciously inserting minus signs into
$V_2$ and $V_3$ for our notational convenience.
(We hope that in context there will be no confusion between
these components of
$\underline{V}$ and the unitary matrix.)
It will be noted that the first two components of $\underline{T}$,
$\underline{U}$, and
$\underline{V}$ give a basis for the six off-diagonal matrices.
However, that $T_3, U_3, V_3$ together with $T, U, V$ and
the unit matrix, must really be
dependent on only 3 independent matrices in the diagonal sector.
We shall see, nevertheless, that this notion is most suited to our
purposes.
Therefore we retain this overspecification and record the
relationships reflecting the degeneracy
for future use.
The notation has been designed for maximum symmetry between
the three spins, and in particular
\beq
T_3 + U_3 + V_3 = 0
\label{15}
\eeq
and
\begin{equation}
T + U + V = 0
\label{16}
\eeq
with our choice of sign conventions.
The remainder of the relationships can be conveniently, but still
redundantly, specified in the form
\begin{eqnarray}
T_3& =& T + 2 V \;, \label{17} \\
U_3 &=& U + 2T \; , \label{18} \\
V_3 &=& V + 2 U \; , \label{19}
\end{eqnarray}
which neatly exposes the cyclical nature of the notation.
In practice, as we shall shortly see, the most immediately useful
relationships are those expressing
the third member of an $SU_2$ set of generators in terms of two
of the singlet operators as in
equations (\ref{17}) -( \ref{19}) above, or their variants utilising
equation (\ref{16}) such as
\beq
V_3 = - V - 2 T \label{20}
\eeq
and
\beq
2U = - T - T_3  \; , \label{21}
\eeq
or again their variants using equation (\ref{17}) such as
\beq
U_3 = 3 U - 2 V_3 \label{22}
\eeq
and
\beq
2U_3 = 3 T - T_3  \;  .
\label{23}
\eeq

We are now ready to work through the details of the $n=3$ case.
There are phase
freedoms
\begin{eqnarray}
                  D & \rightarrow & \exp ( - i \xi T_3)  \exp ( - i
\gamma T) D \; , \label{24} \\
\overline{U}  & \rightarrow & \overline{U} \exp ( i \omega T_3)
\exp \left( \frac{ i \chi}{2} 1
\right) \exp \left( i K T \right)  \; , \label{25}
\end{eqnarray}
and we can take
\beq
V_3 = \exp \left( \frac{ - i \chi}{2} \right) \exp ( - i \rho T_3 )
\exp ( - i \mu T )
K ( \varepsilon, \theta, T ) K ( \delta, \psi, V) K (\lambda, \phi,
u) ,
\label{26}
\eeq
where the matrices $K$ are now $(3 \times 3)$ in an obvious
extension of the previous notation.
Notice that this extension can conveniently be viewed in the form
\beq
V_3 = V_2 \exp ( - i \mu T) K (\delta, \psi, V) K (\lambda, \phi,
U) , \label{27}
\eeq
where $V_2$ has been extended by an extra column and row of
zeros, and that the exponential
term involving $T$ can then be taken to the left through any part
of $V_2$ since T commutes
with all the matrices in $V_2$.
This time things are a little more complicated, and before we can
adjust phases it is necessary to
consider the structure implied whenever two K matrices are
contiguous. In expanded form we see
that
\begin{eqnarray}
K( \varepsilon, \theta, T)
K ( \delta, \psi, V) &=& \exp ( -i \varepsilon T_3) \exp ( 2 i \theta
T_2) \exp ( i \varepsilon T_3)
\nonumber \\
& & \exp ( - i \delta V_3) \exp ( 2 i \psi V_2) \exp ( i \delta V_3)
\label{28}
\end{eqnarray}
when it becomes evident that there is a ``phase matrix'' of $\exp
( i \varepsilon T_3 )  \exp ( - i \delta
V_3)$ appearing between the two ``rotations'' $\exp ( 2 i \theta
T_2)$ and $\exp ( 2 i \psi V_2)$.
However, it is clear from equation (\ref{17}) and from equation
(\ref{20}) that this ``phase matrix''
can be expressed as $\exp \left( i [ \varepsilon  + 2 \delta] T
\right)$
$\exp \left( i [ \delta + 2 \varepsilon ] V \right)$,
so it can be seen that the left hand factor may be commuted to
the left hand end of $V_3$, and
that the right hand factor can be commuted one step to the right
of the $K ( \delta, \psi, V)$ factor
in equation (\ref{28}).
The next step is to examine the structure of $V_3$ farther to the
right of the ``rotation matrix''
part of $K (\delta, \psi,V)$ in equation (\ref{26}).

Noting the extra phase we have just moved to the right this part
of $V$ now becomes
$$
\exp ( i [ \delta + 2 \varepsilon ] V ) \exp ( i \delta V_3) K (
\lambda, \phi, U)
$$
\beq
\begin{array}{rl}
= & \exp ( i [ \delta + 2 \varepsilon ] V ) \exp ( i \delta V_3) \exp
( - i \lambda U_3)  \; \times \\ & \\
   & \exp (2 i \phi U_2 )  exp ( i \lambda U_3) \;\;\; . \end{array}
\label{29}
\eeq
This time it is clear that at least some part of the ``phase matrix''
before the $\phi$ ``rotation
matrix'' can not be moved farther to the right.
However, we can choose to use equations (\ref{19}) and (\ref{22})
to write the first line of equation
(\ref{29}) as
$$
\exp \left( 2 i [ \delta + \varepsilon + \lambda] V_3 \right)
\exp \left(- i U[ 4 \varepsilon + 2 \delta + 3 \lambda ] \right)
$$
revealing that it is possible to move the right hand factor to the
right through the $\phi$ ``rotation
matrix'' and leaving the residual ``phase matrix'' in terms of
$V_3$ alone.
The final step is then to express the (now) three phases to the
right of equation (\ref{26}) in terms
of $T_3$ and $T$.
This is easily seen to have the form
$$
\exp \left( i T    [ \delta + 2 \varepsilon + 3 \lambda ] \right)
\exp \left( i T_3 [ \delta + 2 \varepsilon +     \lambda ] \right)
$$
by using equations (\ref{21}) and (\ref{23}).

We can now adjust the phases in equations (\ref{22}) and (\ref{23}),
so that taking
\begin{eqnarray}
\xi &=& \delta + 2 \varepsilon + \lambda  \; , \label{30} \\
\gamma &=& \delta + 2 \varepsilon + 3 \lambda \; , \label{31} \\
\omega &=& \varepsilon + \rho \; , \label{32} \\
K &=& \mu - \varepsilon - 2 \delta \; , \label{33}
\end{eqnarray}
and calling
\beq
\delta + \varepsilon + \lambda = \Delta \; , \label{34}
\eeq
the form
$$
V_3 =  \exp ( 2 i \theta T_2) \exp ( 2 i \psi V_2) \;\; \times
$$
\beq
\exp ( 2 i \Delta V_3 ) \exp ( 2 i \phi U_2 )  \label{35}
\eeq
emerges as the final four parameter form of the mixing matrix for
the $n=3$ case.

As this is currently thought to be the most important physical
case, we pause here to retrieve
some well known parameterizations before moving on to higher
numbers of generations of
quarks.

The first thing to realize is that the particular grouping of
rotations and phases which we have
presented, although very convenient for counting parameters and
demonstrating the principles,
is by no means unique.
Indeed the very obvious construction
\beq
V_3 = \exp ( 2 i \theta_{23} U_2 )
K ( -  \delta_{13}, - \theta_{13}, V)
\exp ( 2i \theta_{12} T_2 )
\label{36}
\eeq
where the phase remains on both sides of the $1-3$ rotation
matrix in $K$, is precisely
the recommended `Standard Form'' given in reference
\cite{RevPP}
and credited primarily to Chau and Keung \cite{C&K}.
Expanding into matrix form we see that this is
\beq
V_3 = \left[
\begin{array}{lll}
c_{12} c_{13} & s_{12} c_{13} & s_{13} \exp ( - i \delta_{13} ) \\
-s_{12}c_{23}  - c_{12} s_{23}  s_{13} \exp ( i \delta_{13}) &
 c_{12}c_{23}  - s_{12} s_{23} s_{13}   \exp ( i \delta_{13}) &
s_{23} c_{13} \\
s_{12} s_{23}  - c_{12} c_{23} s_{13}  \exp ( i \delta_{13}) &
-c_{12} s_{23}  - s_{12} c_{23} s_{13} \exp ( i \delta_{13}) &
c_{23} c_{13}
\end{array} \right] \,,
\label{37}
\eeq
where $c_{12}$ and $s_{23}$ denote respectively $\cos
\theta_{12}$ and
$\sin \theta_{23}$ etc, as in equation (3) of reference
\cite{RevPP}.
The axis of rotation is indicated by the missing index.

On the other hand, the original $KM$ matrix \cite{K&M} is of
``Euler angle'' type,
involving ``rotations'' about only two axes.
This time we may write this as
\begin{eqnarray}
V_3&=& \exp ( - 2 i \theta_2 U_2)
\exp ( - 2i  \theta_1 T_2)
\exp \left( \frac{i[\delta + \pi]}{3} \right) \times \nonumber \\
& &\exp ( - 2 i [ \delta + \pi] T ) \exp ( 2 i \theta_3 U_2 )
\; , \label{38}
\end{eqnarray}
where the existence of an overall phase (involving $\pi$ which
has the familiar mathematical value
and should not be confused with the four parameters) is needed
in our notation to recover
equation (4) of reference \cite{RevPP}.
This can be expanded as
\beq
V_3 = \left[
\begin{array}{lll}
c_1 & - s_1 c_3 & - s_1 s_3 \\
s_1 c_2 & c_1 c_2 c_3 - s_2 s_3 \exp ( i \delta)  & c_1 c_2 s_3
+ s_2 c_3 \exp ( i \delta ) \\
s_1 s_2 & c_1 s_2 c_3 + c_2 s_3 \exp ( i \delta) & c_1 s_2 s_3
- c_2 c_3 \exp (i \delta )
\end{array} \right] \;,
\label{39} \eeq
where this time $c_1$ denotes $\cos \theta_1$ etc., and the index
on the angles shows the axis
of rotation directly.

It is now clear from the last example that the complex entries in
the $CKM$ matrix can be
contained in four positions.
This raises the amusing possibility of a description in which the
mixing form current to mass
eigenstates exactly contrives to put complex entries only in the
final row and column thus
``interpreting'' the $CP$ violation in the kaon system purely in
terms of intermediate top and
bottom exchange contributions. One way to achieve this is to
take
\beq
V_3 = \exp ( 2 i \theta T_2) K ( - {\scriptstyle \frac{1}{2}} \delta,
- \phi,
V ) \exp ( 2
i \psi T_2) \; , \label{40}
\eeq
so that the expanded form
\beq
V_3 = \left[ \begin{array}{lll}
\begin{array}{l}
 [\cos \theta \cos \phi \cos \psi \\ \; - \sin \theta \sin \psi]
\end{array}&
\begin{array}{l}
[\cos \theta \cos \phi \sin \psi  \\ \; +
\sin \theta \cos \psi]  \end{array}&
\cos \theta \sin \phi \exp ( i \delta) \\ & & \\
\begin{array}{l}[-\sin \theta \cos \phi \cos \psi  \\ \; -  \cos
\theta \sin
\psi] \end{array} &
\begin{array}{l} [\cos \theta \cos \psi  \\ \; - \sin \theta \cos
\phi \sin
\psi] \end{array} &
- \sin \theta \sin \phi \exp ( i \delta) \\ & & \\
- \cos \psi \sin \phi \exp ( - i \delta) &
- \sin \phi \sin \psi \exp (- i \delta ) & \; \cos \phi
\end{array} \right] \; , \label{41}
\eeq
shows this feature directly.

Finally, we turn to the description of the $CKM$ matrix in
terms of phases of the unitarity
triangles \cite{RevPPref}.
Curiously, the key step \cite{{Kayser1},{Kayser2}} in making
the connection is to parameterize
the $CKM$ matrix so that all the complex terms are in the top
left hand corner.
We take the form
\beq
V_3 = \exp (- 2 i \theta T_2)
           \exp ( - 2 i \psi V_2)
           \exp \left( \frac{i \pi}{3} [ 1 - 6 U] \right) \;
           \exp \left( \frac{i \varepsilon}{3} [ 1 - 6 V ] \right)
           \exp ( 2 i \phi T_2 ) \;\;,
\label{42}
\eeq
which expanded out reads
\beq
V_3 = \left[ \begin{array}{lll}
\begin{array}{l}[- \cos \theta \cos \psi \cos \phi  \\ \; + \sin
\theta \sin
\phi \exp ( i \delta)]
\end{array}&
\begin{array}{l}-[ \cos \theta \cos \psi \sin \phi  \\ \; + \sin
\theta \cos
\phi \exp (i \delta)]
\end{array}&
\cos \theta \sin \psi \\ & & \\
\begin{array}{l}[-\sin \theta \cos \psi \cos \phi  \\ \; -  \cos
\theta \sin
\phi \exp (i \delta)]
\end{array} &
\begin{array}{l} [- \sin \theta \cos \psi  \sin \phi  \\ \; + \cos
\theta
\cos \phi \exp (i \delta)]
\end{array} &
\sin \theta \sin \psi \\ & & \\
\;  \sin \psi \cos \phi &
\;  \sin \psi \sin \phi &
\; \cos \psi  \end{array} \right] \; . \label{43}
\eeq

To display the connections to the Kayser
\cite{{Kayser1},{Kayser2}} form more clearly we define
$\lambda$ by
\beq
\cos \psi = \lambda \cos \delta
\; , \label{44}
\eeq
and $r_{ij}$, for $i = u, c$ and $j = d,s,$ by
\beq
r_{ij} \tan \delta = \tan (\arg V_{ij}) \; ,  \label{45}
\eeq
where (as we shall soon see directly) the four $r_{ij}$ are
related by a single constraint.
{}From the top left hand four entries of $V_3$ we now see directly
that
\beq
r_{ud} = \frac{ \tan \theta \tan \phi}{\tan \theta \tan \phi -
\lambda} \; ,\label{46}
\eeq
\beq
r_{cd} = \frac{\tan \phi}{\tan \phi + \lambda \tan \theta}
\; ,\label{47}
\eeq
\beq
r_{us} = \frac{\tan \theta}{\tan \theta + \lambda \tan \phi} \;
,\label{48}
\eeq
and
\beq
r_{cs} = \frac{ 1}{1 - \lambda \tan \theta \tan \phi}\; . \label{49}
\eeq
Eliminating $\lambda, \phi$ and $\theta$ from these equations
reveals
\beq
\frac{(1 - r_{cd}) ( 1 - r_{us})}{r_{cd} \;\;r_{us}} =
\frac{(1 - r_{ud}) ( 1 - r_{cs})}{r_{ud}\;\;r_{cs}}\; , \label{50}
\eeq
as the constraint equation expected.
Then we find
\beq
\lambda^2 = \frac{(1 - r_{ud}) (1 - r_{cs})}{r_{ud} r_{cs}}\; ,
\label{51}
\eeq
with alternative expressions yielded by the use of the constraint
equation.
Reintroducing $\delta$ by equation (\ref{44}) relates $\tan \delta$
to a complicated quotient of
sums of products of the tangents of the angles of the unitarity
triangles.
We do not quote this directly, as we find no simple expression,
although the algebra is direct and
straightforward.
Finally, we can substitute equations (\ref{50}) and (\ref{51}) back
into pairs of equations (\ref{46})
to (\ref{49}) to reveal
\beq
\tan^2 \theta = \frac{(1 - r_{cs}) \; r_{us}}{r_{cs} \; (1 -
r_{us})}\; , \label{52}
\eeq
and
\beq
\tan^2 \phi = \frac{ ( 1 - r_{us}) \; r_{ud}}{r_{us} ( 1 -
r_{ud})} \; , \label{53}
\eeq
where alternative expressions are available by using the constraint
equation (\ref{50}) yet again.
Now that $\lambda$ and $\delta$ are known (at least implicitly),
equation (\ref{44}) gives $\psi$
to complete the connection between our parameterization and the
unitary triangle angles of
Kayser \cite{{Kayser1},{Kayser2}}. We find the algebraic
complexity disappointing, but the
connections are at least clearly made.

We finally treat all cases with 4 or more generations.
Consider expanding from $n$ to $n+1$ where $n \geq 3$.
The first two new matrices introduced will be $\lambda$ matrices
whose indices are $n^2$ and
$n^2 + 1$, and which have entries $1$ and $-i$
respectively in the top right hand corner,
with their conjugates appearing in the bottom left hand corner.
To enable easy visualization we denote these as $2\Sigma_1$ and
$2\Sigma_2$, where the $n$-
dependence has been suppressed.
Clearly $\Sigma_1$ and $\Sigma_2$ are a part of an $SU(2)$ set of
generators,
the third member of which we call $\Sigma_3$ with entries $1/2$
and $-1/2$ in the top left hand
corner and the bottom right hand corner respectively.
Again, consider the new diagonal matrix introduced by the
expansion from $n$ to $n + 1$.
It is, of course, $\lambda_{n ( n + 2)}$.
This has entries $ \left[ \frac{2}{n ( n + 1 )} \right]^{1/2}$ down
the first $n$ diagonal places,
and $(-) \left[ \frac{2 n}{n+1} \right]^{1/2}$ in the final diagonal
place.

Obviously $\lambda_{n(n+2)}$ commutes with the whole set of
$U(n)$ matrices parameterising
$V_n$, and we now write (in an obvious extension of equation
(\ref{27}))
\beq
V_{n+1} = V_n \exp \left( \frac{- i \nu \lambda_{n(n+2)}}{2
\sqrt{3}} \right) K ( \eta, \zeta, \Sigma)
\ldots \; , \label{54}
\eeq
where there are now $n$ new $K$ factors implied, of which we
have shown only the first
explicitly.
As previously, the exponential factor can be moved to the left as
required.
Now the first new $K$ factor can be expanded as before in the
form
\beq
K(\eta, \zeta, \Sigma) = \exp ( - i \eta \Sigma_3) \exp ( 2 i \zeta
\Sigma_2) \exp ( i \eta \Sigma_3)
\; , \label{55}
\eeq
and the now familiar task is to remove the first exponential factor
by expressing it in terms of
diagonal matrices which either commute with everything to the
left or through at least one term
to the right.
Our method is a straightforward extension of that used in the
$n=3$ case.
We introduce a diagonal matrix $C_{n+1}$ which has entries
$\frac{1}{2} \left[ \frac{(n-1)}{3
(n+1)} \right]^{1/2}$ in the top left hand corner and the bottom
right hand corner, and entries
$(-) \left[ \frac{1}{3(n-1)(n+1)} \right]^{1/2}$ in the remaining
$(n-1)$ diagonal places.
This has been designed to be traceless, and to be normalized so
that the trace of its square is the
same as the corresponding matrices in the $3 \times 3$ case.
It is trivial to see that
\beq
\Sigma_3 = \left[ \frac{n}{2 (n+1)} \right]^{1/2} \lambda_{n (
n+2)} + \left[ \frac{ 3 ( n -
1)}{(n + 1)} \right]^{1/2} C_{n+1} \; . \label{56}
\eeq
Obviously $\lambda_{n(n+2)}$ commutes with the entire
structure to the left of $K ( \eta, \zeta,
\Sigma)$ in equation (\ref{54}), and also $C$ commutes with the
$\underline{\Sigma}$ which
appear to its right in equation (\ref{55}).
This latter point is, of course, by construction in analogy with the
$3 \times 3$ case, but is perhaps
intuitively even easier to see now that the matrices are more
sparse.

A last word should probably be said concerning the counting of
parameters in the general case as
displayed by the present analysis.
As we construct $V_{n+1}$, working from the left in equation
(\ref{54}) we first encounter
$V_n$ with $(n-1)^2$ independent real parameters conveniently
viewed as $ \frac{1}{2} n ( n-1)$ angles
and $\frac{1}{2} ( n-1)(n-2)$ phases.
Next we find $n$ factors of $K$, each having the structure
shown in equation (\ref{55}), namely
that of a rotation surrounded by exponential phase factors.
Finally, there is a phase factor carried on the new diagonal matrix
introduced at this level.
What we have shown however is that the phase to the left of the
first new $K$ factor may be
removed by expressing it in terms of a part which commutes to
the left to be absorbed on phases
of $\overline{U}$, and part which commutes one step to the
right.
Finally the phase to the extreme right of the new $K$ factors may
be absorbed into the phases of
$D$.
Thus, overall, there are $n$ new rotation parameters, and $(n-1)$
new phases.
The number of angles then becomes $\frac{1}{2} n ( n-1) + n =
\frac{1}{2} n
(n+1)$, and the number of phases
$\frac{1}{2}(n-1)(n-2) + (n-1) =  \frac{1}{2} n (n-1)$ as previously
stated.
Perhaps it should be emphasized that, just as in the $n=3$ case,
there are many possible variants
of representation of $V_n$ and arising in much the same way.
We do not expand on this theme, however, since currently the
physical interest is in the $n=3$
case and there is no evidence of further generations of quarks and
leptons.
One of the authors (K.J.B.) still retains a hope that a further
generation will be found and that the
economy of orthogonal organising symmetries \cite{KJB} will be
utilised by nature.
In that event, the analysis presented here would be immediately
utility in describing possible mass
breaking schemes.

\bigskip
\noindent{\large{\bf Acknowledgements}}

We are pleased to thank Professor B Kayser for early access to his
work.
We are grateful for partial financial support from PPARC.

\end{document}